\documentclass[aps,prd,superscriptaddress,groupedaddress,nofootinbib]{revtex4}  
\pdfoutput=1 

\usepackage[T1]{fontenc} 

\usepackage{graphicx}  
\usepackage{dcolumn}   
\usepackage{bm}        
\usepackage{amssymb}   
\usepackage{amsmath}
\usepackage{ulem}
\usepackage{color}
\usepackage[colorlinks]{hyperref}
\usepackage{setspace}
\usepackage[font=small,labelfont=bf]{caption}
\usepackage{wrapfig} 

\newcommand{\be}{\begin{equation}}
\newcommand{\ee}{\end{equation}}
\newcommand{\bea}{\begin{eqnarray}}
\newcommand{\eea}{\end{eqnarray}}
\newcommand{\nn}{\nonumber}

\def\x{{\bf x}}
\def\k{{\bf k}}
\def\p{{\bf p}}
\def\y{{\bf y}}
\def\r{{\bf r}}

\begin{document}

\title{Cosmological bounds on TeV-scale physics and beyond}

\author{Niayesh Afshordi}
\email{nafshordi@pitp.ca}
\affiliation{Department of Physics \& Astronomy, University of Waterloo, Waterloo, ON, N2L 3G1, Canada}
\affiliation{Perimeter Institute for Theoretical Physics, 31 Caroline St. N., Waterloo, ON, N2L 2Y5, Canada}
\author{Elliot Nelson}
\email{enelson@pitp.ca}
\affiliation{Perimeter Institute for Theoretical Physics, 31 Caroline St. N., Waterloo, ON, N2L 2Y5, Canada}

\date{\today}

\begin{abstract}
We study the influence of {\it the fluctuations} of a Lorentz invariant and conserved vacuum on cosmological metric perturbations, and show that they generically blow up in the IR.  We compute this effect using 
the K\"{a}ll\'{e}n-Lehmann spectral representation of stress correlators in generic quantum field theories, as well as the holographic bound on their entanglement entropy, both leading to an IR cut-off that scales as the fifth power of the highest UV scale (in Planck units). One may view this as analogous to the Heisenberg uncertainty principle, which is imposed on the phase space of gravitational theories by the Einstein constraint equations. The leading effect on cosmological observables come from anisotropic vacuum stresses which imply: i)  any extension of the standard model of particle physics can only have masses (or resonances) $\lesssim 24$ TeV, {\it and} ii) perturbative quantum field theory {\it or} quantum gravity become strongly coupled beyond a UV scale of $ \Lambda \lesssim 1$ PeV. Such a low strong coupling scale is independently motivated by the Higgs hierarchy problem.
This result, which we dub the {\it cosmological non-constant problem}, can be viewed as an extension of the cosmological constant (CC) problem, demonstrating the non-trivial UV-IR coupling and (yet another) limitation of effective field theory in gravity. However, it is more severe than the old CC problem, as vacuum fluctuations cannot be tuned to cancel due to the positivity of spectral densities or entropy. We thus predict that future advances in  cosmological observations {\it and} collider technology will sandwich from above and below, and eventually discover, new (non-perturbative) physics beyond the Standard Model within the TeV-PeV energy range.
\end{abstract}

\maketitle


\section{Introduction}
The cosmological constant problem is arguably one of the deepest and most long-standing puzzles in theoretical physics \cite{Straumann:2002tv, Weinberg:1988cp}.
Naively one would expect energy in the vacuum to couple to the spacetime metric and drive accelerated expansion at an enormous rate, governed by a UV scale $\Lambda$ above which there are is no coupling to gravity from modes in the vacuum.
The observation of a large and very slowly accelerating universe \cite{Riess:2004nr,Perlmutter:1998np} raises the question of why the vacuum does not gravitate in this way (or why $\Lambda \sim $  meV is so small, while quantum field theory is well tested up to energies 15 orders of magnitude higher), and much effort has been invested in understanding this apparent gross inconsistency \cite{Burgess:2013ara,Aslanbeigi:2011si,Narimani:2014zha}.
At this level, the cosmological constant problem asks how the expectation value of the vacuum stress-energy tensor $\langle T^{(V)}_{\mu\nu}\rangle$ affects the metric $g_{\mu\nu}$ as a source in Einstein's equation.
Quantum mechanically, the vacuum is characterized by higher order correlators, beginning with the two-point function $\langle T^{(V)}_{\mu\nu}(x) T^{(V)}_{\mu\nu}(y)\rangle$. Just as the vacuum expectation value $\langle T^{(V)}_{\mu\nu}\rangle$ can influence the homogeneous background cosmology, these higher-order correlations influence higher-order classical statistics of (inhomogeneous) cosmological perturbations.
Thus, there is a {\it Cosmological non-Constant} (or CnC) problem which asks how the ultraviolet physics of the vacuum encoded in these correlators affects spacetime geometry on infrared, cosmological scales.

In particular, we will here address the question of how two-point functions for metric perturbations can be affected by the vacuum, and whether cosmological observations can place constraints on UV physics.  While we find that energy and momentum density fluctuations are small in the infrared, the vacuum stress components $T_{ij}^{(V)}$ scale as
\be\label{rhonoisepower}
\langle T_{ij}^{(V)}(\k) T_{kl}^{(V)} (\k') \rangle \sim \delta^3(\k+\k')\Lambda^5,
\ee
leading to point-like correlation in real space,
\be\label{rhoPoisson}
\langle T_{ij}^{(V)}(\x)T_{kl}^{(V)}(\y) \rangle \sim \delta^3(\x-\y) \Lambda^5.
\ee
The gravitational potential $\Phi$ is affected by anisotropic stress as $k^2\Phi\sim M_{\rm p}^{-2} A^{ij}T_{ij}$, where $A^{ij}$ is a tensor which picks out the anisotropic part of $T_{ij}$, and $M_{\rm p}\equiv(8\pi G)^{-1/2}$ is the reduced Planck mass. So Eq. \eqref{rhonoisepower} yields a contribution
\be\label{Delta_vac}
\big(\Delta^{(V)}_{\Phi}\big)^2\sim \frac{\Lambda^5}{M_p^4 k}
\ee
to the dimensionless power spectrum $\Delta_\Phi^2$, defined by
\be
\langle\Phi_{\k}\Phi_{\k'}\rangle \equiv (2\pi)^3 \delta^3(\k+\k')\Delta_{\Phi}^2(k) \left(k^3/2\pi^2\right)^{-1}.
\ee
These fluctuations will affect observable quantities. To a first-order approximation, we can require that $\Delta_\Phi^2\lesssim1$ to maintain consistency with the observation of a homogeneous background cosmology. Requiring that $\Delta^{(V)}_{\Phi}\lesssim1$ on the largest accessible scales, $k\sim H_0$, leads to the condition
\be
\Lambda \lesssim (M_{\rm p}^4 H_0)^{1/5} \approx 2 \ \rm PeV.
\ee 
In the following sections we will make this schematic statement more precise, and strengthen the bound by studying how the UV scale can be constrained with cosmological information about the amplitude of $\Phi$.

Vacuum stress-energy fluctuations - in particular two-point correlators - have been studied in a flat space context \cite{Ford:2005sp} as well as on curved space \cite{Osborn:1999az,Cho:2014ira}, including de Sitter space \cite{Wu:2006ew,PerezNadal:2009hr} and inflation \cite{Hsiang:2011ix}. The physical effects of stress tensor fluctuations on the gravitational field were reviewed in \cite{Ford:2007gb}. We also note that our results - in particular Section \ref{sec:scalarfield} - share some overlap with results in stochastic gravity (see \cite{Hu:2008rga} and references therein).

Stress-energy correlators generically feature UV divergences, requiring a method of regularization or renormalization to obtain a finite and physically meaningful result. We will encounter this obstacle in our treatment below, and note that regularization of stress-energy correlators has been studied extensively in the literature (e.g., see \cite{Frob:2013sxa,Hack:2012qf} and references therein).

In Section \ref{sec:einsteinfrw}, we present Einstein's constraint equations for scalar and vector perturbations in a flat FRW background, sourced by stress-energy perturbations. We do not consider tensor perturbations in this work, as they do not appear in linearized Einstein constraint equations, and thus are much  less sensitive to the short-wavelength vacuum fluctuations, due to time averaging over the past light cone. For further discussion of this point see Section \ref{sec:conclusion}.

In Section \ref{sec:kl}, we parametrize vacuum stress-energy in a general quantum field theory context using the K\"{a}ll\'{e}n-Lehmann spectral representation, and identify the large-scale behavior of metric perturbations.
In Section \ref{sec:kSZ}, we study the influence of vacuum stress-energy on gravitational forces on astrophysical objects and the Hubble law. In Section \ref{sec:poisson}, we introduce a classical toy model for a conserved, generally covariant two-point function for stress-energy, featuring a lack of spatial correlation, or Poisson statistics. We compute the influence of this matter source on cosmological perturbations and on the cosmic microwave background (CMB) through the integrated Sachs-Wolfe (ISW) effect. In Section \ref{sec:scalarfield}, we study vacuum stress-energy for a massive scalar field and find a spectral density closely connected to that of the Poisson model of Section \ref{sec:poisson}, through analytic continuation.
Finally, in Section \ref{sec:entropybound} we present an independent constraint on UV physics from requiring that the entropy of a region of space not exceed the holographic bound given by the Bekenstein-Hawking area law.
We discuss our results and conclude in Section \ref{sec:conclusion}.

Those interested in our results may skip to the ends of sections \ref{sec:kSZ}, \ref{sec:poisson}, and \ref{sec:entropybound}, and note the boxed equations there.
We will use the $(-+++)$ metric signature, and natural units $\hbar=c=k_B=1$ throughout. We will denote the scalar product of four-momentum as $k_\mu k^\mu\equiv k_4^2$ to distinguish it from the magnitude of three-momentum $|\k|\equiv k$.

\section{Metric Perturbations in FRW Background}
\label{sec:einsteinfrw}

We work in the conformal Newtonian gauge for metric perturbations,
\be
ds^2 = a^2(\eta)\left[-(1+2\phi)d\eta^2 +  2V_i dx_i d\eta + (1-2\psi)d\mathbf{x}^2\right].
\ee
The Einstein equations for scalar metric perturbations are, in Fourier space,
\bea
3\mathcal{H}(\mathcal{H}\phi+\psi')+k^2\psi &=& 4\pi Ga^2\delta T_0^0, \hspace{2cm} \label{00einstein} \\
-ik_i(\mathcal{H}\phi+\psi') &=& 4\pi Ga^2\delta T_i^0, \label{i0einstein} \\
\Big(\psi'' + \mathcal{H}(2\psi+\phi)' + (2\mathcal{H}' + \mathcal{H}^2)\phi - \frac{1}{2}k^2(\phi-\psi)\Big)\delta_{ij} + \frac{1}{2}k_ik_j(\phi-\psi) &=& 4\pi Ga^2\delta T_j^i, \label{ijeinstein}
\eea
where $0$ components and $'$ derivatives refer to conformal time $\eta$. Eqs. \eqref{00einstein} - \eqref{ijeinstein} can be solved for $\psi$ and $\phi$ in terms of the stress-energy,
\bea
-k^2\psi &=& 4\pi G \left(\delta T_{00}-\frac{3\mathcal{H}}{k^2}ik^i\delta T_{i0}\right), \label{psieinstein} \\
-k^2\phi &=& 4\pi G\left(\delta T_{00} - \frac{3\mathcal{H}}{k^2} ik^i\delta T_{i0} + \left(\delta^{ij}-3\frac{k^ik^j}{k^2}\right)\delta T_{ij}\right), \ \ \ \ \ \ \ \label{phi}
\eea
as well as the time derivatives
\bea\label{psideriv}
-k^2\psi' &=& 4\pi G \left[ - \mathcal{H} \delta T_{00} + \left(1 + \frac{3\mathcal{H}^2}{k^2}\right)ik^i\delta T_{i0} - \mathcal{H}\left(\delta^{ij}-3\frac{k^ik^j}{k^2}\right)\delta T_{ij}\right]. \\
-k^2\phi' &=& 4\pi G\bigg[-\mathcal{H}\delta T_{00} + \left(1 + \frac{3\mathcal{H}^2}{k^2}\right)ik^i\delta T_{i0} + \left(\delta^{ij}-3\frac{k^ik^j}{k^2}\right)\left(\delta T_{ij}' - \mathcal{H} \delta T_{ij}\right)\bigg]. \label{phideriv}
\eea
Note that in the subhorizon regime $k\gg\mathcal{H}$, at leading order we can drop $\mathcal{O}(\mathcal{H})$ and $\mathcal{O}(\mathcal{H}^2)$ terms.
\bea
-k^2\psi &=& 4\pi G \delta T_{00}, \label{psisubhorizon} \\
-k^2\phi &=& 4\pi G\big[\delta T_{00} + (\delta^{ij}-3\hat{k}^i\hat{k}^j)\delta T_{ij}\big], \label{phisubhorizon} \\
-k^2\psi' &=& 4\pi G ik^i\delta T_{i0}, \label{psiderivsubhorizon} \\
-k^2\phi' &=& 4\pi G\big[ik^i\delta T_{i0} + (\delta^{ij}-3\hat{k}^i\hat{k}^j)(\delta T_{ij})'\big], \label{phiderivsubhorizon}
\eea
where $\hat{k}_i\equiv k_i/k$.

For vector metric perturbations, Einstein's equations take the form
\be\label{vectorconstraint}
k^2 V_i = 16\pi G (\delta_{ij}-\hat{k}_i\hat{k}_j) \delta T_{j0},
\ee
where the projection $(\delta_{ij}-\hat{k}_i\hat{k}_j)$ picks out the purely vector part of $\delta T_{j0}$. 
The conservation of stress-energy, $\nabla_{\mu}T^{\mu\nu} = 0$, gives us the Fourier space constraints \bea\label{conservationconstraints}
(\delta T_0^0)' &=& i k^i \delta T_i^0 - 3\mathcal{H}\delta T_0^0 + \mathcal{H}\delta T_i^i, \label{T00conserved} \\ 
(\delta T_i^0)' &=& -4\mathcal{H}\delta T_i^0 -i k^j \delta T_i^j,  \label{T0iconserved}
\eea
from which we see that
\be\label{V'}
k^2 V'_i = 16\pi G (\delta_{ij}-\hat{k}_i\hat{k}_j)(ik^l\delta T_{jl} - 2\mathcal{H}\delta T_{j0}). 
\ee
Eqs. \eqref{psieinstein}-\eqref{phideriv}, \eqref{vectorconstraint}, \eqref{V'} show the impact of vacuum stress-energy on metric perturbations. In subsequent sections we will consider specific sources $\delta T_{\mu\nu}$. 

At the linear level, the Einstein equations hold unambiguously as equations for Heisenberg operators, with metric fluctuations treated quantum mechanically in the same way as stress-energy.


The effect of vacuum stress-energy on metric fluctuations which we compute is exact for linear perturbations around a Minkowski spacetime background. Additional matter, or any form of background stress-energy, will lead to curvature in the geometry. In this case, the effect of vacuum stress-energy on the metric backreacts in turn on other (non-vacuum) forms of stress-energy. A fully self-consistent analysis would have to account for this in solving Einstein's equations. The flat-space effect, however, is the leading contribution in an expanding universe, with subleading contributions suppressed by $\mathcal{H}/k$ on sub-Hubble scales. In the super-Hubble regime $k\lesssim\mathcal{H}$, these corrections become large, and the Minkowski space result is no longer the approximate solution.

\section{Vacuum stress-energy in the K\"{a}ll\'{e}n-Lehmann representation}
\label{sec:kl}

We now move on to consider stress-energy fluctuations in a general quantum field theory context. We can parametrize the non- time-ordered two point function for a conserved, Lorentz invariant stress-energy source in the K\"{a}ll\'{e}n-Lehmann spectral representation \cite{WeinbergQFT} as
\bea\label{twopointKL}
&&\langle T_{\mu\nu}(x)T_{\alpha\beta}(y)\rangle_{c,s} \equiv 
\langle T_{\mu\nu}(x)T_{\alpha\beta}(y)\rangle_s - \langle T_{\mu\nu}(x)\rangle\langle T_{\alpha\beta}(y)\rangle
\nonumber\\
&&=\int\frac{d^4k}{(2\pi)^4}e^{ik\cdot(x-y)}\int_0^{\infty}d\mu\left[\rho_0(\mu)P_{\mu\nu}P_{\alpha\beta} + \rho_2(\mu)\left(\frac{1}{2}P_{\mu\alpha}P_{\nu\beta}+\frac{1}{2}P_{\mu\beta}P_{\nu\alpha}-\frac{1}{3}P_{\mu\nu}P_{\alpha\beta}\right)\right]2\pi\delta(k_4^2+\mu),\nonumber\\
\eea
where the projection tensors $P_{\mu\nu}\equiv\eta_{\mu\nu}-k_{\mu}k_{\nu}/k_4^2$ ensure stress-energy conservation $\partial_{\mu}T^{\mu\nu}=0$. Here, the $c$ subscript denotes the connected part of the two-point function; in what follows we will omit this subscript and implicitly discuss only the connected part.
 $\langle\rangle_s$ denotes the symmetrized two-point correlation, which makes it the expectation value of a Hermitian operator, and thus a potential quantum observable. In Appendix \ref{app:KLrep}, we discuss the tensor structure and spectral densities $\rho_{0,2}$, and show how scalar contractions of Eq. \eqref{twopointKL} relate to the spectral densities. The subscripts refer to spin-0 or spin-2 states; as shown in Appendix \ref{app:KLrep}, only the tensor structure for the spin-0 part contributes to correlators with the stress-energy trace $T^\mu_\mu$; the spin-2 tensor structure comes from the traceless part of $T_{\mu\nu}$. Eq. \eqref{twopointKL} is consistent with unitarity if and only if $\rho_{0,2}(\mu)\geq0$.\footnote{\label{KLfootnote}This may be shown by requiring that the expectation value $\langle T(J)^2\rangle$ of the square of $T(J)\equiv\int d^4x T_{\mu\nu}J^{\mu\nu}$ be non-negative for an arbitrary real-valued source function $J^{\mu\nu}(x)$. We thank Andrew Tolley for correspondence on this point.}

It is important to note that for UV physics, only high frequencies will contribute to $\langle T_{\mu\nu}(x)T_{\alpha\beta}(y)\rangle$; the spectral densities $\rho_{0,2}$ vanish for $\mu$ smaller than the masses of heavy particles.

Fourier transforming the spatial arguments and defining $T_{\mu\nu}(\mathbf{k})\equiv\int d^3\mathbf{x} T_{\mu\nu}(\mathbf{x})e^{-i\mathbf{k}\cdot\mathbf{x}}$, we have at equal times
\be\label{TmunuKL}
P^{(V)}_{\mu\nu\alpha\beta}(\k,t) = \int_0^\infty\frac{d\mu}{2\sqrt{\mathbf{k}^2+\mu}}\left[\rho_0(\mu)P_{\mu\nu}P_{\alpha\beta} + \rho_2(\mu)\left(\frac{1}{2}P_{\mu\alpha}P_{\nu\beta}+\frac{1}{2}P_{\mu\beta}P_{\nu\alpha}-\frac{1}{3}P_{\mu\nu}P_{\alpha\beta}\right)\right],
\ee
where $-k_4^2=\mu$ in $P_{\mu\nu}$, and the equal-time power spectra are defined by
\be
\langle T^{(V)}_{\mu\nu}(\mathbf{k},t)T^{(V)}_{\alpha\beta}(\mathbf{k}',t)\rangle \equiv (2\pi)^3\delta^3(\mathbf{k}+\mathbf{k}') P^{(V)}_{\mu\nu\alpha\beta}(\k,t).
\ee

Since we are interested in the leading effect on IR scales $k\equiv|\k|$ from physics at UV scales $\mu$, we neglect $k$ in these expressions; that is, we drop terms at higher orders in $k^2/\mu$. We define
\be
J^{(s)}_m(\mu_{\rm max})\equiv\frac{1}{2}\int_0^{\mu_{\rm max}} d\mu \mu^{m} \rho_s(\mu), \ \ s=0,2. \label{J_def}
\ee
We will drop the argument for $\mu_{\rm max}\rightarrow\infty$. Note that $J_m$ has mass dimension $2m+6$, and in particular $J_{-1/2}$, which is most relevant here, is dimension five.

Due to the projection tensors in Eq. \eqref{twopointKL}, which impose stress-energy conservation, long-wavelength vacuum pressure and stress fluctuations are large on infrared scales, and energy and momentum density fluctuations are suppressed in comparison. This is because $0$ indices lead to a suppression by $k/\sqrt{\mu}$, so that only spatial components $\langle T_{ij}T_{kl}\rangle$ are dominant. 
At leading order, then, 
\be
P^{(V)}_{ijkl}(\k,t) = \delta_{ij}\delta_{kl} J^{(0)}_{-1/2} + \left(\frac{1}{2}\delta_{ik}\delta_{jl}+\frac{1}{2}\delta_{il}\delta_{jk} - \frac{1}{3}\delta_{ij}\delta_{kl}\right)J^{(2)}_{-1/2} + \mathcal{O}(k^2), \label{spatialTT}
\ee
and components of $\langle T_{\mu\nu}T_{\alpha\beta}\rangle$ with fewer spatial indices are suppressed by powers of $k$.

\subsection{Cosmological Influence of Vacuum Stress-Energy}
\label{sec:klcosmo}
We now treat the stress-energy of Eq. \eqref{twopointKL} as a perturbation $\delta T^{(V)}_{\mu\nu}$ in Einstein's equations, and find the contribution to metric (equal-time) two-point functions sourced by the stress-energy as given in Eq. \eqref{twopointKL}. We will place a $V$ superscript on power spectra to indicate that only the contribution to two-point functions sourced by the vacuum are shown, not the full two-point functions. The ``$0$'' index will indicate conformal time $\eta$. 

In Einstein's equations, \eqref{phi}-\eqref{phideriv}, the traceless tensor $(\delta^{ij}-3\hat{k}^i\hat{k}^j)$ picks out only anisotropic stress to source metric perturbations, and the transverse projection for vector modes in Eq. \eqref{V'} has the same effect. Thus, (isotropic) pressure from the vacuum does \textit{not} couple to metric perturbations, and as noted above, other components of $T_{\mu\nu}$ are small in the infrared, leaving only anisotropic stress as the vacuum source of infrared geometry.
We consider the subhorizon regime $k\gg\mathcal{H}$. Going from a Minkowski to FRW geometry\footnote{We also expect the curvature to introduce contributions scaling as $\mathcal{H}/k$, which would likely prove difficult to compute, since we would be computing vacuum expectation values in curved space, where there is no unique vacuum. In Section \ref{sec:conclusion} we comment on the case of (Anti) de Sitter space, where we may be able to compute vacuum stress-energy correlators.} introduces a scale factor, since $k^3P_\phi(k)$ is a physical quantity (it sets the real-space variance of $\phi$) and should therefore depend only on the physical wavenumber, $\sim1/k_{\rm ph}=a/k$. The same applies to $k^3P_{\phi'}(k)/a^2$, with an additional $a^2$ coming from converting physical time $d/dt$ to comoving time $d/d\eta$.

Using Eq. \eqref{spatialTT}
with Eqs. \eqref{phisubhorizon} and \eqref{phiderivsubhorizon}, then, we find
\hspace{-1.3cm}
\begin{flalign}
k^3 P^{(V)}_{\phi}(k,a) &= \frac{3}{2} \frac{J^{(2)}_{-1/2}}{M_{\rm p}^4}\frac{a}{k} = \frac{3a}{4M_{\rm p}^4 k} \int \frac{d\mu}{\sqrt{\mu}} \rho_2(\mu)   , \label{p_phi} \\
\frac{1}{a^2}k^3 P^{(V)}_{\phi'}(k,a) &= \frac{3}{2} \frac{J^{(2)}_{1/2}}{M_{\rm p}^4}\frac{a}{k} = \frac{3a}{4M_{\rm p}^4 k} \int d\mu \sqrt{\mu} \rho_2(\mu) \label{p_phi_prime}
\end{flalign}
with $P^{(V)}_{\psi}$ and $P^{(V)}_{V,ij}(k)$ suppressed by $\mathcal{O}(k/\sqrt{\mu})$. In the second line, the time derivative acting on Eq. \eqref{twopointKL} simply adds a factor of $k_0^2\approx -k_4^2=\mu$ in the integral.
Here, $M_{\rm p}\equiv(8\pi G)^{-1/2}$ is the reduced Planck mass. 

In the following section, we will use this result to constrain the spectral density $\rho_2$ via its effect on gravitational forces $\sim\nabla\phi$ on astrophysical objects. Although only high frequencies contribute to the integrals in Eqs. \eqref{p_phi}-\eqref{p_phi_prime}, we argue in Section \ref{sec:scalarfield} below that for a finite range of time, transforming stress-energy correlators to real time will lead to correlations at large time separation. Furthermore, two high frequency modes contributing to $\phi$ can generate at nonlinear order a low frequency mode. We consider a resulting cosmological effect in the following section.

\section{Constraints from the Hubble flow} 
\label{sec:kSZ}

Here we consider the gravitational force $\nabla\phi(\x)$ on galaxy clusters in the kinematic Sunyaev-Zel'dovich effect, and identify constraints on vacuum stress-energy via its contribution to this force.

Take the action for a relativistic point particle,
\be
S_p = -m \int dt \sqrt{1+2\phi({\bf x},t)-|\dot{\bf x}|^2 },
\ee
where, following the discussion above, we neglect other metric perturbations compared to $\phi$, so that $-g_{00} = 1+2\phi$. Furthermore, the proper time of an observer at the origin is given by:
\be
\tau = \int dt \sqrt{1+2\phi(0,t)}.
\ee
Now, expressing the particle action to second order in $\phi$ and $ \dot{\bf x}$, in terms of the observer's proper time, yields
\be
S_p \simeq m\int d\tau \left[-1 + \frac{1}{2} |\dot{\bf x}|^2+\phi(0,t)-\phi({\bf x},t)+\frac{1}{2}\phi({\bf x},t)^2-\frac{3}{2}\phi(0,t)^2+\phi({\bf x},t)\phi(0,t) \right].
\ee
Given that $\phi$ is oscillating at high frequencies (given by the UV scale of the theory), the linear terms cannot affect the dynamics of macroscopic observables (e.g. a galaxy). In fact, averaging over the oscillations, the only non-vanishing and non-constant contribution to the effective  Newtonian potential is the cross term:
\be
\Phi_N({\bf x},t) \simeq  -\langle \phi({\bf x},t)\phi(0,t) \rangle_{\rm high},
\ee
where we have taken the quantum expectation value only over high frequencies. Averaging over low frequencies gives
\be
\langle \Phi_N({\bf x},t) \rangle_{\rm low} = -\int \frac{d^3k}{(2\pi)^3} \exp(i {\bf k\cdot x}) P_{\phi}(k),
\ee
where $P_\phi(k) \propto k^{-4}$ is given by Eq. (\ref{p_phi}). While the integral in  $\Phi_N$ (or the $\phi$-correlation function) is formally divergent, the acceleration is finite and given by: 
\be
\langle {\bf x}\cdot \nabla \Phi_N({\bf x},t) \rangle \simeq  -\int \frac{d^3k}{(2\pi)^3} (i {\bf k\cdot x}) \exp(i {\bf k\cdot x}) P_{\phi}(k) = \frac{3}{16\pi} \frac{J^{(2)}_{-1/2}}{M^4_p} |{\bf x}|. \label{pert_force}
\ee

The surprising conclusion is that, at second order in perturbations, the high frequency oscillations lead to a uniform and constant centripetal acceleration. While Eq. (\ref{pert_force}) provides the perturbation to the Newtonian gravitational force due to high frequency fluctuations, its integral over the classical trajectory provides the change in radial velocity. Within linear perturbation theory, this leads to a constant infall peculiar velocity, or a negative offset to the Hubble law:\footnote{Observers elsewhere would see the same effect in their proper time, which depends differently on $\phi(\x)$.}

\be
{\bf v} = H {\bf x} - \frac{g(\Omega_m)}{H} \nabla\Phi_N = H {\bf x} - \frac{3g(\Omega_m)}{16\pi H}  \frac{J^{(2)}_{-1/2}}{M^4_p} \frac{{\bf x}}{|{\bf x}|}, \label{monopole} 
\ee
where $g(\Omega_m)$ comes from solving for radial trajectories, to linear order in perturbation theory in a $\Lambda$CDM background:
\be
g(\Omega_m) \equiv  \frac{\int_0^1 (1-a^2) h(a,\Omega_m)^{-3} a^{-1} da }{\int_0^1  h(a,\Omega_m)^{-3} a^{-1} da},~ h(a,\Omega_m) \equiv \sqrt{\Omega_m a^{-3} +1- \Omega_m}
\ee

By measuring the kinematic Sunyaev-Zel'dovich effect of X-ray selected galaxy clusters, the Planck satellite \cite{Ade:2013opi} has measured the mean radial peculiar velocity to be $\langle v_r \rangle = 72 \pm 60$ km/s, which implies  $\langle v_r \rangle > -25$ km/s $= - 8.3 \times 10^{-5} c$ at the 95\% confidence level. 
Plugging this into Eq. (\ref{monopole}), and using the definition of $J^{(2)}_{-1/2}$ (Eq. \ref{J_def}), as well as $\Omega_m = 0.3$ yields:
\be\label{kSZconstraint}
\boxed{ \left[J^{(2)}_{-1/2} \right]^{1/5} = \left[\frac{1}{2} \int \frac{d\mu}{\sqrt{\mu}} \rho_2(\mu) \right]^{1/5} < 1.0~ {\rm PeV}, }
\ee
at the 95\% confidence level. 

Given that $\rho_2(\mu) \sim \mu^2$ at high energies for a weakly coupled quantum field theory (e.g., see Eq. \ref{scalarfieldrho} below), we can interpret this upper limit as an upper limit on the energy scale of perturbative quantum field theory, or quantum gravity.  Beyond this scale, the theory must be strongly coupled (with no weakly coupled local UV completion), as $\rho_2(\mu)$ must decay faster than $\mu^{-1/2}$ for the $J^{(2)}_{-1/2}$ integral to converge.

\section{A Poisson Sprinkling of Phase Space}
\label{sec:poisson}

In this section, we consider a toy model for the vacuum comprised of a classical collection of massive particles in phase space, which is manifestly Lorentz invariant and conserved. We will see that a Lorentz invariant stress-energy correlator describes this model, along with a flat Poisson spectrum, as in Eq. \eqref{rhonoisepower}, and will find an ISW plateau for masses $\gtrsim 24 \ \rm TeV$, which is already constrained by CMB observations. We shall show in Section \ref{sec:scalarfield} below that a free massive scalar quantum field theory reproduces an identical structure to this toy model on large scales. Here and in the following subsection we restrict to Minkowski space; in Section \ref{poissoncosmology} we generalize to an FRW background.

With a phase-space distribution $f$, the stress-energy tensor integrates over momenta with the Lorentz invariant measure
\be\label{Tmunu}
T^{\mu\nu}(\mathbf{x},t)=\int\frac{d^3\mathbf{p}}{p^0}p^{\mu}p^{\nu}f(\mathbf{x},\mathbf{p},t),
\ee
Here, $p^\mu$ is the four-momentum for classical point particles of mass $m$.
We consider a distribution $f(\mathbf{x},\mathbf{p},t)$ characterized (like Eq. \eqref{rhoPoisson}) by point-like correlation or Poisson noise\footnote{For zero correlation between different phase space points as in Eq. \eqref{Poissonphasespace}, the probability for finding $N$ particles in a given region of phase space is a Poisson distribution.} in phase space,
\be\label{Poissonphasespace}
\langle f(\mathbf{x},\mathbf{p},t)
f(\mathbf{x}',\mathbf{p}',t)\rangle
- \langle f(\mathbf{x},\mathbf{p},t)\rangle\langle 
f(\mathbf{x}',\mathbf{p}',t)\rangle=\langle f(\mathbf{x},\mathbf{p},t)\rangle \delta^3(\mathbf{x}-\mathbf{x}')\delta^3(\mathbf{p}-\mathbf{p}').
\ee
The expectation value here is taken with respect to an ensemble of distributions of particles, from which a particular realization $f$ is drawn. From now on we will ignore the constant $\langle f\rangle^2$ term, as this will only shift the background, and correlators for stress-energy fluctuations. We take the  mean distribution to be uniform in phase space\footnote{To keep the distribution normalizable, we assume that at high momenta $p\gg m$, $\langle f(p) \rangle$ falls to zero.},
\be
\langle f(\mathbf{x},\mathbf{p},t)\rangle = \langle f(t)\rangle.
\ee
We will see below that this results in a Lorentz invariant two-point function. 

To evaluate the stress-energy two-point function, we note first that the phase space distribution is related to the initial distribution $f_0$ at time $t_0$ by
\be\label{fevolution}
f(\mathbf{x}_,\mathbf{p},t) =
f_0\left(\mathbf{x}-(t-t_0)\frac{d\mathbf{x}}{dt},\mathbf{p},t_0\right),
\ee
Putting together Eqs. \eqref{Poissonphasespace} and \eqref{fevolution}, we find
\be\label{poissontwopointmomentum}
\langle T^{\mu\nu}(\mathbf{x},t)T^{\alpha\beta}(\mathbf{x}+\Delta\mathbf{x},t+\Delta t)\rangle_c = \int\frac{d^3\mathbf{p}}{(p^0)^2}p^{\mu}p^{\nu}p^{\alpha}p^{\beta}
\langle f_0\rangle
\delta^3\left(\Delta x^i - \Delta t p^i / p^0\right). \ \ \ 
\ee
For spacelike separation $|\Delta\x|>\Delta t$, the delta function vanishes for all $\p$. For timelike separation it fixes
\be\label{pfixed}
p^{\mu}=m\frac{\Delta x^{\mu}}{\sqrt{-\Delta x_{\gamma}\Delta x^{\gamma}}},
\ee
and contributes a normalization factor $m^3\Delta t^2(-\Delta x_{\gamma}\Delta x^{\gamma})^{-3/2}$. That is, the only particles in the original collection contributing to the correlation function at separation $\Delta x^\mu$ are those with the corresponding four-momentum given by Eq. \eqref{pfixed}.
Putting everything together, we have
\be\label{poissontwopoint}
\langle T^{\mu\nu}(x)T^{\alpha\beta}(x+\Delta x)\rangle_c
= m^5 \langle f_0\rangle\frac{\Delta x^{\mu}\Delta x^{\nu}\Delta x^{\alpha}\Delta x^{\beta}}{(-\Delta x_{\gamma}\Delta x^{\gamma})^{7/2}}\Theta(-\Delta x_{\gamma}\Delta x^{\gamma}).
\ee
Eq. \eqref{poissontwopoint} is a conserved, generally covariant two-point function in Minkowski space. We will use it as a proxy for vacuum stress-energy, and argue below that the two behave similarly on large scales.

\subsection{Spectral Density for Poisson Stress-Energy}
\label{sec:rhopoisson}

The spectral densities for Poisson stress-energy in the K\"{a}ll\'{e}n-Lehmann representation can be found by transforming to Fourier space. Defining $T_{\mu\nu}(k)\equiv\int d^4x e^{-ik\cdot x} T_{\mu\nu}(x)$, we have
\be
\langle T_{\mu\nu}(k)T_{\alpha\beta}(k') \rangle = (2\pi)^4\delta^4(k+k')\int d^4x e^{ik\cdot x} \langle T_{\mu\nu}(y)T_{\alpha\beta}(x+y)\rangle.
\ee
Using Eq. \eqref{projectrho0} to relate the left hand side to $\rho_0(-k_4^2)$, we find
\be
2\pi \rho_0^{\rm Poisson}(-k_4^2) = S_0^{\mu\nu\alpha\beta} \int d^4x e^{ik\cdot x} \langle T_{\mu\nu}(y)T_{\alpha\beta}(x+y)\rangle_{c,s}.
\ee
For Poisson stress-energy the Fourier transform can be most easily obtained from Eq. \eqref{poissontwopointmomentum}. Eliminating the spatial integral with the delta function there, and integrating over time, we find
\be\label{poissonpintegral}
\rho_0^{\rm Poisson}(-k_4^2) = \frac{1}{9}\langle f_0\rangle m^4\int\frac{d^3p}{p_0}\delta(k_{\mu}p^{\mu}) = \frac{1}{18}\langle f_0\rangle m^4\int d^4p \delta(p^2+m^2) \delta(k_{\mu}p^{\mu}) \Theta(p_0).
\ee
For $k_4^2<0$, $\rho_0^{\rm Poisson}$ vanishes due to the delta function since $p^2<0$. For $k_4^2>0$, we can evaluate the integral in a Lorentz invariant manner by replacing the delta functions with integrals over phases, and using the stationary phase approximation:
\be
\int d^4p \delta(p^2+m^2) \delta(k_{\mu}p^{\mu})\Theta(p_0) = \int_{-\infty}^{\infty} \frac{dxdy}{(2\pi)^2} \int_{p_0>0} d^4p e^{i(p^2+m^2)x+i(k_{\mu}p^{\mu})y}.
\ee
The phase is stationary with respect to $p$ at $2xp_\mu+yk_\mu=0$, so
\bea
\int_{p_0>0} d^4p \delta(p^2+m^2) \delta(k_{\mu}p^{\mu}) &\simeq& \sqrt{\frac{1}{(-i)^3i}} \int \frac{dxdy}{(2\pi)^2} \frac{\pi^2}{2x^2} \exp\left[i\left(-\frac{k_4^2}{4x}y^2+m^2x\right)\right] \nn \\
&=& \sqrt{\frac{1}{(-i)^3i^2}} \frac{\sqrt{\pi}}{8} \int \frac{dx}{x^2} \sqrt{\frac{4x}{k_4^2}} e^{im^2x} = \sqrt{\frac{1}{(-i)^3i^2}} \frac{\sqrt{\pi}}{4}\sqrt{\frac{m^2}{k_4^2}} \int \frac{dz}{z^{3/2}} e^{iz},\nonumber\\ \label{Gammaintegral}
\eea
where in the first line the $\pi^2/2x^2$ comes from evaluating the $p$ integrals after shifting $p\rightarrow -ky/2x$ and extracting the phase, and in the second line we have integrated over $y$. The integral $\int dz e^{iz}/z^{3/2}$ is divergent. However, we can deform the contour to integrate along down the positive imaginary axis, around the origin, and back up with a small positive real part, so that $\int dz e^{iz}/z^{3/2} \rightarrow \frac{2}{\sqrt{i}} \int_0^\infty dz e^{-z}/z^{3/2}=\Gamma(-1/2)/\sqrt{i}=-2\sqrt{\pi/i}$. The appearance of factors of $\sqrt{i}$ reveal an ambiguity in our regularization scheme; we choose the prescription which gives positive spectral densities (see footnote \ref{KLfootnote}).

Also recall that since the Poisson stress-energy correlator is totally symmetric in its indices, its tensor structure in the K\"{a}ll\'{e}n-Lehmann representation will be totally symmetric, $P_{\mu\nu}P_{\alpha\beta} + P_{\mu\alpha}P_{\nu\beta} + P_{\mu\beta}P_{\nu\alpha}$, which fixes
\be\label{Poissonrho0rho2}
\rho_2^{\rm Poisson}=(6/5)\rho_0^{\rm Poisson}.
\ee
Putting everything together, then, we have
\be\label{Poissonrho2}
\rho_2^{\rm Poisson}(-k_4^2) = \frac{\pi}{30}\langle f_0\rangle m^4\sqrt{\frac{m^2}{k_4^2}}\Theta(k_4^2).
\ee
In the following section we will make use of this result to study the cosmological influence of Poisson stress-energy.

The regularization of the divergence at small $x$ or $z$ in Eq. \eqref{Gammaintegral} controls the divergent behavior of Poisson stress-energy at large four-momentum $p$, which corresponds to approaching the light cone $\Delta x^2=0$, as seen in Eq. \eqref{pfixed}. This gives us a well-behaved result for power spectra which, matches the IR limit of quantum field theory stress-energy correlators in the K\"{a}ll\'{e}n-Lehmann representation, as we will see in section \ref{sec:scalarfield}.

\subsection{Cosmological Influence of Poisson Stress-Energy}
\label{poissoncosmology}

We can now find the contribution to metric two-point functions sourced by the stress-energy as given in Eq. \eqref{poissontwopoint}, which we take to be a small perturbation $\delta T^{(\rm P)}_{\mu\nu}$.
We will place a $(\rm P)$ superscript on power spectra to indicate that only the contribution to two-point functions sourced by Poisson stress-energy is shown, not the full two-point functions.
Making use of the symmetrical index structure of Poisson stress-energy, which leads to Eq. \eqref{Poissonrho0rho2}, we have
\be
\langle \delta T^{(\rm P)}_{\mu\nu}(k) \delta T^{(\rm P)}_{\alpha\beta}(k')\rangle_c = (2\pi)^4 \delta^4(k+k') 2\pi \rho_2^{\rm Poisson}(-k_4^2) \frac{1}{2}\left(P_{\mu\nu}P_{\alpha\beta}+P_{\mu\alpha}P_{\nu\beta}+P_{\mu\beta}P_{\nu\alpha}\right).
\ee
For the Poisson spectral density, Eq. \eqref{Poissonrho2}, we find from Einstein's equations in the $k\equiv|\k|\gg\mathcal{H}$ regime the following results for two-point correlators in momentum space, where we define $\langle X(\mathbf{k},\omega)X(\mathbf{k}',\omega')\rangle\equiv (2\pi)^4\delta^3(\k+\k')\delta(\omega+\omega')P_X(k,\omega)$,
\bea\label{phi'phi'}
P^{(\rm P)}_{\phi'}(k,\omega)&=&\frac{\pi^2}{10}\frac{a m^5}{M_{\rm p}^4}\langle f_0\rangle\frac{\omega^2}{k^4}\frac{1}{\sqrt{k^2-\omega^2}}\left(1-\frac{4}{3}\frac{k^2}{k^2-\omega^2}\right) \Theta(k^2-\omega^2), \\
P^{(\rm P)}_{\psi'}(k,\omega)&=&\frac{\pi^2}{40}\frac{a m^5}{M_{\rm p}^4}\langle f_0\rangle \frac{\omega^2}{(k^2-\omega^2)^{5/2}} \Theta(k^2-\omega^2), \\
P^{(\rm P)}_{V',ij}(k,\omega)&=&\frac{2\pi^2}{15}\frac{a m^5}{M_{\rm p}^4}\langle f_0\rangle(\delta_{ij}-\hat{k}_i\hat{k}_j)\frac{-\omega^2}{k^2}\frac{1}{(k^2-\omega^2)^{3/2}} \Theta(k^2-\omega^2). \label{V'V'}
\eea
We have included the scale factor $a(\eta)$ to account for the effect of FRW space, through the following argument: While the power spectrum $P_{\phi'}(k,\omega)$ is not a physical quantity itself, the combination $k^3 \omega P_{\phi'}(k,\omega)/a^2$ is a physical quantity, since the real-space correlation $\langle\dot{\phi}(\x_1,t_1)\dot{\phi}(\x_2,t_2)\rangle$ for a small range of momenta $(\Delta\k,\Delta\omega)$ peaked around  $(\k,\omega)$ goes like this combination. Thus, $k^3 \omega P_{\phi'}(k,\omega)/a^2$ should depend only on physical frequencies and wavenumbers, $k_{\rm ph}=k/a$ and $\omega_{\rm ph}=\omega/a$. Making this replacement in Eqs. \eqref{phi'phi'}-\eqref{V'V'} adds the factor of $a(\eta)$, as shown above. A time-dependence to the frequency spectrum is valid for sub-Hubble times $\omega\gtrsim H(\eta)$.

Upon integrating over frequencies $\omega$ to obtain correlators in real time, we obtain divergent pieces going as powers of $\omega_{\rm max}$, but for $P^{(\rm P)}_{\phi'}$ and $P^{(\rm P)}_{V',ij}$ a finite, positive part remains (despite the negativity of $P^{(\rm P)}_{V',ij}$), which we take to be the physical effect of vacuum stress-energy. We also note that as emphasized in section \ref{sec:klcosmo} this effect comes from the correlator of spatial components $\langle\delta T_{ij}\delta T_{kl}\rangle$, and from anisotropic stress, that is from $\rho_2$ rather than the isotropic $\rho_0$ part of Eq. \eqref{twopointKL}.

\subsection{ISW Effect from Poisson Stress-Energy.} \label{sec:PoissonISW}

We now show that Poisson stress-energy contributes a flat spectrum (independent of $l$) to the CMB angular power spectrum, contributing most significantly at high $l$ through the ISW effect. Including the influence of vector modes \cite{Boubekeur:2009uk}, the ISW effect is
\be
\frac{\delta T^{\rm ISW}(\hat{\r})}{T}=\int_{\eta_{LSS}}^{\eta_{\rm today}} d\eta(\phi' + \psi' + V'_i\hat{r}^i),
\ee
Here, the integration is carried out over the past light cone, that is $\phi'=\phi'(r,\hat{\r},\eta)$ with $r(\eta)/r_{LSS}\approx 1-\eta/\eta_{\rm today}$.
The contribution to the CMB angular power spectrum is well approximated by the Limber approximation \cite{LoVerde:2008re}, which, for integration over the light cone, gives
\bea\label{C_l_ISW}
C_l^{\rm ISW} &\simeq& \int \frac{d\Omega}{4\pi}\frac{dr}{r^2}\int\frac{d\omega}{2\pi}\left[P^{(\rm P)}_{\phi'}\left(\sqrt{k_\perp^2+\omega^2},\omega\right)
 + \hat{r}^i\hat{r}^jP^{(\rm P)}_{V',ij}\left(\sqrt{k_\perp^2+\omega^2},\omega\right)\right], \hspace{1.25cm}
 \eea
where $k_\perp = (l+\frac{1}{2})/r$. Here, we have dropped the $\langle\psi'\psi'\rangle$ power spectrum and cross-spectrum $\langle\phi'\psi'\rangle$, which only have power-law divergent contributions after integration over $\omega$, as well as the cross-spectra $\langle\phi'V'_i\rangle$ and $\langle \psi'V'_i\rangle$, which vanish due to the transverse nature of the vector modes. We have also dropped power-law divergent contributions to the $\langle\phi'\phi'\rangle$ and $\langle V'_iV'_j\rangle$ correlators. Furthermore, for the vector term, only the isotropic part of Eq. \eqref{V'V'} contributes.

The power spectra in Eqs. \eqref{phi'phi'} and \eqref{V'V'} go like $1/k_\perp^2$ after integration over $\omega$, canceling the $r$-dependence in the integrand of Eq. \eqref{C_l_ISW}. However, the scale factor in Eqs. \eqref{phi'phi'}-\eqref{V'V'} leads to the radial integral contributing $\int dr a(r_{LSS}-r)\approx\int d\eta a(\eta)=\int dt\approx H_0^{-1}$ for $\Omega_m=0.3$ and $\Omega_\Lambda=0.7$.
Making use of Eqs. \eqref{phi'phi'} and \eqref{V'V'} and using $(\Omega_m,\Omega_\Lambda)=(0.3,0.7)$, we find that at high $l$
\be\label{ISW_Poisson}
(\Delta_l^2)^{\rm ISW}\equiv\frac{l(l+1)C^{\rm ISW}_l}{2\pi}= \frac{49\pi}{720} \frac{m^5 t_0}{M_{\rm p}^4} \langle f_0 \rangle \simeq \frac{49\pi}{720} \frac{m^5}{M_{\rm p}^4 H_0} \langle f_0\rangle,
\ee
which is a flat spectrum with $l$, and thus largest compared to CMB anisotropies in linear theory at the end of the damping tail at high $l$. In Appendix \ref{SPT}, we show that at the $95\%$ confidence level, current CMB observations roughly translate to:
\be
(\Delta_l^2)^{\rm ISW}\simeq\frac{2.2 \ \mu K^2}{(2.73 K)^2} \lesssim  3.0\times10^{-13}.
\ee
Since $M_{\rm p} = 2.44 \times 10^{18} ~{\rm GeV}$ and $H_0\approx 70 \ \rm km/s/Mpc = 1.5\times 10^{-33} \ \rm eV$,
requiring the ISW contribution, Eq. \eqref{ISW_Poisson}, to be consistent with the observed values implies
\be\label{Poissonconstraint}
\boxed{m \lesssim (24~{\rm TeV})\left[(2\pi)^{3} \langle  f_0 \rangle \over 2\right]^{-1/5}.}
\ee
In Section \ref{sec:scalarfield}, we motivate this choice for $\langle f_0\rangle$ by drawing an analogy to quantum field theory of a free massive scalar field. 

\section{Poisson-like Stress-energy for a Massive Scalar Field}
\label{sec:scalarfield}

In this section, we compute the stress-energy of a massive scalar field, and show that it behaves similarly to the Poisson spectrum considered above. While we find that due to the requirement of positive energy, the amplitude of stress-energy fluctuations vanishes for small $|k^2|$, for large enough time separations $\Delta t$ the spectral density at high energies contributes to the stress-energy correlator in the same way as the low-energy Poisson spectral density.

For a massive scalar field we have
\be
T_{\mu\nu}(x)=\partial_{\mu}\varphi\partial_{\nu}\varphi - \frac{1}{2}g_{\mu\nu}\left[(\partial\varphi)^2+m^2\varphi^2\right].
\ee
Defining the momentum space quantities
\be
\varphi_k\equiv\int d^4x e^{-ik\cdot x}\varphi(x), \ \ \ T_{\mu\nu}(k)\equiv\int d^4x e^{-ik\cdot x} T_{\mu\nu}(x),
\ee
we have
\be
T_{\mu\nu}(k)=\int\frac{d^4p}{(2\pi)^4}M_{\mu\nu}(p,k-p)\varphi_p\varphi_{k-p},
\ee
where
\be
M_{\mu\nu}(k,k')\equiv -k_{\mu}k'_{\nu} + \frac{1}{2}g_{\mu\nu}(k^{\alpha}k'_{\alpha} - m^2).
\ee
The two-point correlation is then given by
\be
\langle T_{\mu\nu}(k)T_{\alpha\beta}(k')\rangle = \int\frac{d^4p}{(2\pi)^4}\frac{d^4p'}{(2\pi)^4} M_{\mu\nu}(p,k-p)M_{\alpha\beta}(p',k'-p')\langle\varphi_p\varphi_{k-p}\varphi_{p'}\varphi_{k'-p'}\rangle.
\ee
The scalar field two-point correlation can be evaluated in terms of commutation relations for creation and annihilation operators:
\be
\langle\varphi_k\varphi_{k'}\rangle = (2\pi)^4 \delta^4(k+k') 2\pi \delta(k^2+m^2)\Theta(k_0).
\ee
Evaluating the connected four-point correlation in the same way, we find
\bea
\langle\varphi_p\varphi_{k-p}\varphi_{p'}\varphi_{k'-p'}\rangle_c &=& (2\pi)^4\delta^4(k+k')\delta(p^2+m^2)(2\pi)^2\delta((k-p)^2+m^2) \nn \\
&&\times[(2\pi)^4\delta^4(p+p')+(2\pi)^4\delta^4(k-p+k')]\Theta(p_0)\Theta(k_0-p_0).
\eea
Consequently, letting $p\rightarrow-p$ in the integral, we have
\bea
\langle T_{\mu\nu}(k)T_{\alpha\beta}(k')\rangle_c &=& \delta^4(k+k')\int d^4p M_{\mu\nu}(p,k-p)M_{\alpha\beta}(p,k-p) \nn \\
&& \hspace{1cm} \times (2\pi)^2 \delta(p^2+m^2) \delta((k-p)^2+m^2) \Theta(p_0)\Theta(k_0-p_0).
\eea
To find $\rho_2$ via Eq. \eqref{projectrho2}, we contract with $S_2^{\mu\nu\alpha\beta}$,
\bea
&&S_2^{\mu\nu\alpha\beta} M_{\mu\nu}(p,k-p)M_{\alpha\beta}(p,k-p) = \frac{1}{60}\left[-k^4+2m^2k^2+8m^4+(2k^2+4m^2)p\cdot k+2(p\cdot k)^2\right]. \nn \\
\eea
Here we have used the fact that $p^2=-m^2$.
We also have
\be\label{spin0contraction}
S_0^{\mu\nu\alpha\beta} M_{\mu\nu}(p,k-p)M_{\alpha\beta}(p,k-p) = \frac{1}{9} \left[k^2 - m^2 - p\cdot k \right]^2.
\ee
For timelike $k$ we are free to choose $\k=0$, so $k\cdot p=-k_0p_0$, and
\bea
2\rho_2(-k^2)\Theta(k_0) &=& \int \frac{d^4p}{(2\pi)^4} (2\pi)\delta(p^2+m^2) \delta((k-p)^2+m^2) \Theta(p_0)\Theta(k_0-p_0) \left[S_2^{\mu\nu\alpha\beta} M_{\mu\nu} M_{\alpha\beta}\right] \nn \\ 
&=& \frac{1}{(2\pi)^3} \int\frac{d^3\p}{2\omega_p}\delta(-k_0^2+2k_0\omega_p)\Theta(k_0-\omega_p)\left[S_2^{\mu\nu\alpha\beta} M_{\mu\nu} M_{\alpha\beta}\right] \nn \\ &=& \frac{1}{(2\pi)^3} \int\frac{d^3\p}{2\omega_p}\frac{\omega_p}{2k_0|\p|}\delta\big(|\p|-\sqrt{k_0^2/4-m^2}\big)\Theta(k_0 - 2m)\frac{k_0^4}{120}\left[1-4\frac{m^2}{k_0^2}\right]^2,\nonumber\\ 
\eea
where $\omega_p\equiv\sqrt{|\p|^2+m^2}$. In the second line, the delta function sets the step function equal to unity, but requires the addition of the step function in the third line. The factor of $2\Theta(k_0)$ indicates the unsymmetrized expectation value. Symmetrizing as in Eq. \eqref{projectrho2}, and replacing $k_0^2\rightarrow-k_4^2$ to write the expression in its covariant form, we have
\be
\rho_2(-k_4^2) = \frac{1}{240}\frac{k_4^4}{16\pi^2}\sqrt{1+4\frac{m^2}{k_4^2}}\left[1+4\frac{m^2}{k_4^2}\right]^2 \Theta(-k_4^2-4m^2). \label{scalarfieldrho}
\ee
(For spacelike $k$ on the other hand, we are free to set $k_0=0$, leading to $\Theta(p_0)\Theta(-p_0)=0$ and a vanishing correlator, so that Eq. \eqref{scalarfieldrho} applies for all $k$.)
In the same way, we find from Eq. \eqref{spin0contraction} that
\be
\rho_0(-k_4^2) = \frac{1}{18}\frac{k_4^4}{16\pi^2}\sqrt{1+4\frac{m^2}{k_4^2}}\left[\frac{1}{2}-\frac{m^2}{k_4^2}\right]^2 \Theta(-k_4^2-4m^2). \label{scalarfieldrho0}
\ee
Eqs. \eqref{scalarfieldrho}-\eqref{scalarfieldrho0} are consistent with existing literature on scalar-field stress-energy; see e.g. \cite{Hu:2008rga,Martin:2000dda}.

\begin{figure}
\begin{center}
\includegraphics[scale=0.35]{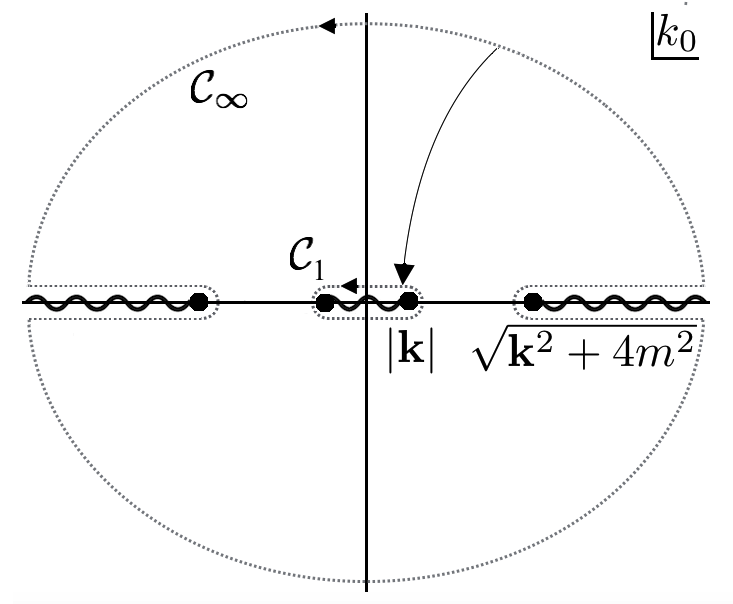}
\caption{The integrations over $\mathcal{C}_\infty$ and $\mathcal{C}_1$ are equivalent.}
\label{figure2}
\end{center}
\end{figure}

In order to find stress-energy correlators in real time at a separation $\Delta t$, we integrate over frequencies $k_0$, as in Eqs. \eqref{twopointKL}-\eqref{TmunuKL}. Physically, in the absence of very energetic particles we do not expect linear macroscopic observables to be sensitive to very high frequency fluctuations in the metric. If we lack access to energies above some scale $\omega_{\rm max}$, the stress-energy correlators that are physically relevant for metric perturbations will be approximately given by a Fourier transform over frequencies that is damped above $\omega_{\rm max}$
\be\label{Jrealtime}
J_\varphi ({\bf k},\Delta t) \equiv \int_{-\infty}^\infty dk_0 \rho_2(k_0^2-\k^2) e^{-ik_0\Delta t} e^{-k_0^2/\omega_{\rm max}^2}.
\ee
For $|k_4^2| \gg m^2$, $J_\varphi$ becomes a gaussian integral, which can be evaluated analytically, and is  
 equal to a polynomial $ \times \exp\left(-\frac{1}{4}\omega_{\rm max}^2\Delta t^2\right)$.
This allows for the integration over $k_0$, which runs over $|k_0|>\sqrt{\k^2+4m^2}$, to be done along the contour $\mathcal{C}_{\infty}$ shown in Figure \ref{figure2}, as long as it is closed at $|k_4^2| \gg m^2$. The parts of the contour around the branch cuts along the real axis give us four times the original integral, while the integration at $|k_0|\rightarrow\infty$ can be dropped for $\omega_{\rm max}|\Delta t| \gg 1$. We can then shrink the contour to $\mathcal{C}_1$, which encloses the branch points at $k_0=\pm|\k|$. Taking $\omega_{\rm max}\gg|\k|$ and dropping terms suppressed by $|\k|^2/m^2$ in Eq. \eqref{scalarfieldrho}, this is
\be
J_\varphi({\bf k},\Delta t)= \frac{m^4}{120\pi^2} \int_{\mathcal{C}_1} dk_0 \sqrt{\frac{m^2}{-k_0^2+\k^2}}e^{-ik_0\Delta t}, ~{\rm~for~} |\Delta t| \gg \omega^{-1}_{\rm max}. \ee
So the scalar-field stress-energy contributes in the same way as a spectral density at low frequencies
\be\label{effectiverho}
\rho_{2,\varphi}^{\rm effective}(-k_4^2) = - \frac{m^4}{120\pi^2}\sqrt{\frac{m^2}{k_4^2}}\Theta(k_4^2),
\ee
integrated for real $k_0$.

The scalar field spectral density, Eq. \eqref{scalarfieldrho}, therefore contributes to correlations in real time in the same way as the effective spectral density, Eq. \eqref{effectiverho}. This is equivalent to the low-frequency spectral density for the Poisson stress-energy of Section \ref{sec:rhopoisson}, so we expect the same stress-energy correlations at long times $\Delta t$ and large distances $|\k|\ll m$ which led to the TeV scale bound discussed there. (The Poisson stress-energy correlation is that of a collection of pointlike particles, which we only ought to compare to the quantum field theory case on spatial scales much larger than the Compton wavelength $m^{-1}$ of the particles.)
 
Comparing to Eq. \eqref{effectiverho}, we see that
\be
|\rho_{2,\varphi}^{\rm effective}| = \frac{2}{(2\pi)^3\langle f_0\rangle} \rho_2^{\rm Poisson}.
\ee
Motivated by the similar behavior of these spectral densities, we consider equating them to fix $\langle f_0\rangle=2/(2\pi)^3$. As noted in Section \ref{sec:PoissonISW}, this places the bound  at $m\lesssim 24 \ \rm TeV$.  We note that the ratio of the effective spectral densities $\rho_0$ and $\rho_2$ for a scalar field is 
equal to the ratio for Poisson stress-energy, strengthening the analogy. Following the above procedure for Eq. \eqref{scalarfieldrho0} to pick out the leading term in the $k_4^2\ll m^2$ limit leads to
\be\label{effectiverho0}
\rho_{0,\varphi}^{\rm effective}(-k_4^2) = \frac{5}{6}\rho_{2,\varphi}^{\rm effective}(-k_4^2).
\ee
We conjecture that the same Poisson behavior arises generically in the $k_4^2\rightarrow0$ limit for massive quantum field theories.

\section{Holographic Entanglement Entropy Bound and an IR Cut-off of Gravity}
\label{sec:entropybound}

In this section, we argue that an IR cut-off in quantum gravity, similar to what we have discovered in the CnC problem above, should have also been anticipated from a holographic bound on entanglement entropy of non-gravitational degrees of freedom. This independent line of reasoning reinforces the interpretation of the scale where metric fluctuations become large (due to stress-energy sources) as a minimum energy scale or maximum distance scale for the theory. It also shows the uniqueness of our results to gravity in comparison to other gauge theories such as QED (see section \ref{sec:EM}), which lack a holographic entropy bound.

Imagine a quantum system, described by the Hamiltonian $H_0$ in its ground state $|0 \rangle_\circ$, whose energy is set to zero for convenience. Now, if we turn on a perturbation $H_{\rm int}$, then the static wavefunction $|0 \rangle_\circ$ is no longer a solution of the Schrodinger equation, but rather will slowly evolve in the Hilbert space of $H_0$ with the amplitudes that are (to first order) given by
\be
\langle n| 0 \rangle_\circ \simeq - \frac{ \langle n | H_{\rm int} | 0 \rangle }{E_n},
\ee  
where $| n\rangle$ for $n>0$ are the excited states of $H_0$. Even though, at any time,  $|0 \rangle_\circ$ is still a pure state,  we can define a density matrix for observation at random times (or averaged over time):
\be
\rho_{\rm int} = \sum_n |\langle n| 0 \rangle_\circ|^2 |n\rangle\langle n| = \sum_n \frac{ |\langle n | H_{\rm int} | 0 \rangle|^2 }{E^2_n} |n\rangle\langle n|,
\ee
which, e.g. for a two-level system, yields a von Neumann entropy
\be 
S_{qubit}= - tr (\rho_{\rm int} \ln \rho_{\rm int} ) \simeq \alpha \left[1-\ln(\alpha)\right] +{\cal O}(\alpha^2),
\ee
where 
\be
\alpha \equiv \frac{ |\langle 1 | H_{\rm int} | 0 \rangle|^2 }{E^2_1},
\ee
can be interpreted as the {\it fine structure constant} for the interaction. 

Now, let us apply this result to the ground state of a quantum field theory (QFT), as we turn on gravity. Of course, one may wonder whether you could start in the global ground state of QFT+gravity. However,  it is not clear whether such a state even exists, and even if it does, our evolving universe is certainly not in its global ground state. Alternatively, we may consider this entropy to be the entanglement entropy of the mixed state of QFT in the (pure) ground state of QFT+gravity, which has a similar parametric dependence on $\alpha$.

Therefore, starting from the QFT ground state, we expect a minimum (entanglement) entropy of $-\alpha_G \ln(\alpha_G)$ for every qubit of the QFT, where $\alpha_G \sim E^2/M_p^2$ is the gravitational fine structure constant. The minimum number of qubits in a volume, up to an energy scale $\Lambda$, e.g. for a Dirac field is given by
\be
\# = 2 \times 2 \times Volume \times \int^\Lambda \frac{d^3 k}{(2\pi)^3} =  \frac{2 \Lambda^3}{3\pi^2} \times Volume ,    
\ee   
where the factors of $2$ account for spins up and down, as well as particles and anti-particles. The condition that this entropy should not exceed the holographic (Bekenstein-Hawking) entropy implies:
\be
S_{BH} = 2\pi M^2_p \times Area > S  =  \# \times \alpha_G\left[1- \ln(\alpha_G)\right] \sim \frac{2 \Lambda^5\left[1+\ln(M^2_p/\Lambda^2)\right] }{3\pi^2 M^2_p} \times Volume. 
\ee
For a spherical region of radius $R$, this yields:
\be
R \lesssim R_{\rm max} \sim \frac{3\pi^3 M^4_p}{\Lambda^5 \left[1+\ln(M^2_p/\Lambda^2)\right] },
\ee
which, as in previous sections, provides an IR cut-off (or scale of strong coupling) in terms of the UV cut-off of the QFT, when gravity is turned on:  
\be
\Lambda_{IR} \sim \frac{\pi}{R_{\rm max}} \sim \frac{\Lambda^5 \left[1+\ln(M^2_p/\Lambda^2)\right]}{3\pi^2 M^4_p}.
\ee
Requiring that our observable universe fits within this IR cut-off yields:
\bea
\Lambda_{IR} < H_0 \simeq 9.5 \times 10^{-33} {\rm eV} \nonumber\\
\Rightarrow \boxed{\Lambda \lesssim 2.4 ~{\rm PeV}. }
\eea

While this is not quite as strong as previous bounds, we should note that it does not use any precision cosmology data. In fact, one could have obtained this upper limit, nearly a century ago, upon the discovery of quantum mechanics and cosmic expansion!

\section{Discussion} 
\label{sec:conclusion}

The nonlinearity of general relativity couples modes on very different scales. Classically, cosmological metric and matter density perturbations evolve nonlinearly and become coupled on very different scales, leading to nonlinear structure formation.
Quantum mechanically, metric perturbations on cosmological scales can also be influenced by short-scale vacuum modes, allowing UV physics to influence classical geometry in the IR. 

\subsection{Summary}

At large distances, we saw that stress-energy fluctuations $\delta T_{\mu\nu}$ are largest from the spatial components and behave like Poisson noise, being uncorrelated in real space (or, having a flat spectrum in $\k$-space, as in Eq. \eqref{spatialTT}). Using these fluctuations as a source for metric perturbations, we found constraints on vacuum stress-energy from measurements of galaxy cluster peculiar velocities (Eq. \eqref{kSZconstraint}), which probe the gravitational potential. Requiring that metric perturbations remain perturbative on the largest accessible scales leads to a similar constraint from Eq. \eqref{p_phi}. These constraints place bounds on the integrated stress-energy spectral density, $\int d\mu\rho_2(\mu)/\sqrt{\mu}$, given explicitly in terms of the stress-energy tensor in Appendix \ref{app:KLrep}. We can interpret this as an upper bound on an energy cut-off beyond which quantum field theory, or quantum gravity, is strongly coupled.
For the Poisson stress-energy correlator,  given in Eq. \eqref{poissontwopoint}, which resembles the stress correlators of a massive scalar quantum field theory in the IR, we found even stronger constraints on the UV scale through the ISW contribution to the CMB (Eq. \eqref{Poissonconstraint}). 

 The proximity of our upper bound of $24$ TeV-$1$ PeV to the electroweak scale suggests a connection to a possible resolution to the Higgs hierarchy problem, which independently requires new physics close to the TeV scale. In particular, the CnC problem points to a gravitational resolution to the hierarchy problem such as (supersymmetric) large extra dimensions with low scale quantum gravity (e.g., \cite{Antoniadis:1998ig,Burgess:2004kd}). The other possibility is a transition to strongly coupled non-perturbative dynamics, such as technicolor \cite{Hill:2002ap}, or theories with emergent Lorentz symmetry \cite{Bednik:2013nxa}. We thus predict that future advances in collider technology and cosmological observations will sandwich, and eventually discover new  TeV scale physics.

\subsection{Directions for Future Work}

Several extensions of our work are possible. A complete account of the influence of the vacuum would include backreaction: matter will move on geodesics now modified by the vacuum-influenced geometry, leading to a different matter stress-energy which will in turn affect the geometry. We leave a full self-consistent analysis to future work. In particular, the influence of the vacuum on geometry and matter geodesics on the scales of gravitationally bound systems would be worth investigating.

While we have focused on the impact of vacuum stress-energy on infrared cosmological modes, we have not considered the effect of modes in the super-Hubble regime $k\ll\mathcal{H}$. In this regime the leading effect of metric perturbations is a shift to the locally observed spatial curvature \cite{Geshnizjani:2005ce}. Our results here suggest that metric fluctuations become large in the IR, $\Delta^2_\phi\sim1/k$, possibly leading to the breakdown of FRW as a background spacetime on very large scales. Given a UV scale which sets the amplitude of vacuum stress-energy, Einstein's equations determine an IR scale at which metric fluctuations sourced by vacuum stress-energy become large. It would be interesting to seek a nonperturbative understanding of the influence of vacuum stress-energy in the $k\rightarrow0$ limit, and the consequential effect on the background geometry in Hubble-sized volumes.

In particular, the IR divergences encountered here may be well-behaved in de Sitter or Anti de Sitter space, with the curvature radius controlling the IR behavior, and a unique vacuum fixed by the symmetry of the spacetime, making curved-space QFT calculations possible. It would be interesting to carry out the same exercise of finding vacuum stress-energy correlators, and their effect on fluctuations in the geometry, in (A)dS, with an exact computation valid on super-curvature scales. The strong coupling on large scales may be dual to a confinement in the boundary QFT.

We have also refrained from discussing the conversion of quantum fluctuations of the metric (superpositions of different gravitational field configurations) into smooth, classical perturbations. In order for this to take place, a mechanism of decoherence is needed to effectively measure the spacetime geometry. We expect that an environment of particles with some characteristic energy will be sensitive to metric fluctuations up to frequencies of the same order, and will select out a classical configuration for metric perturbations with lower frequencies. Energies in the primordial universe, during a reheating or hot big bang era, will reach a maximum before redshifting, and will thus be most sensitive to vacuum stress-energy during this epoch. It would be interesting to apply our results in this context, which would require a better understanding of stress-energy as a source for metric perturbations on super-Hubble modes, as noted above.

\subsection{Renormalization}

\begin{figure}
\begin{center}
\includegraphics[scale=0.2]{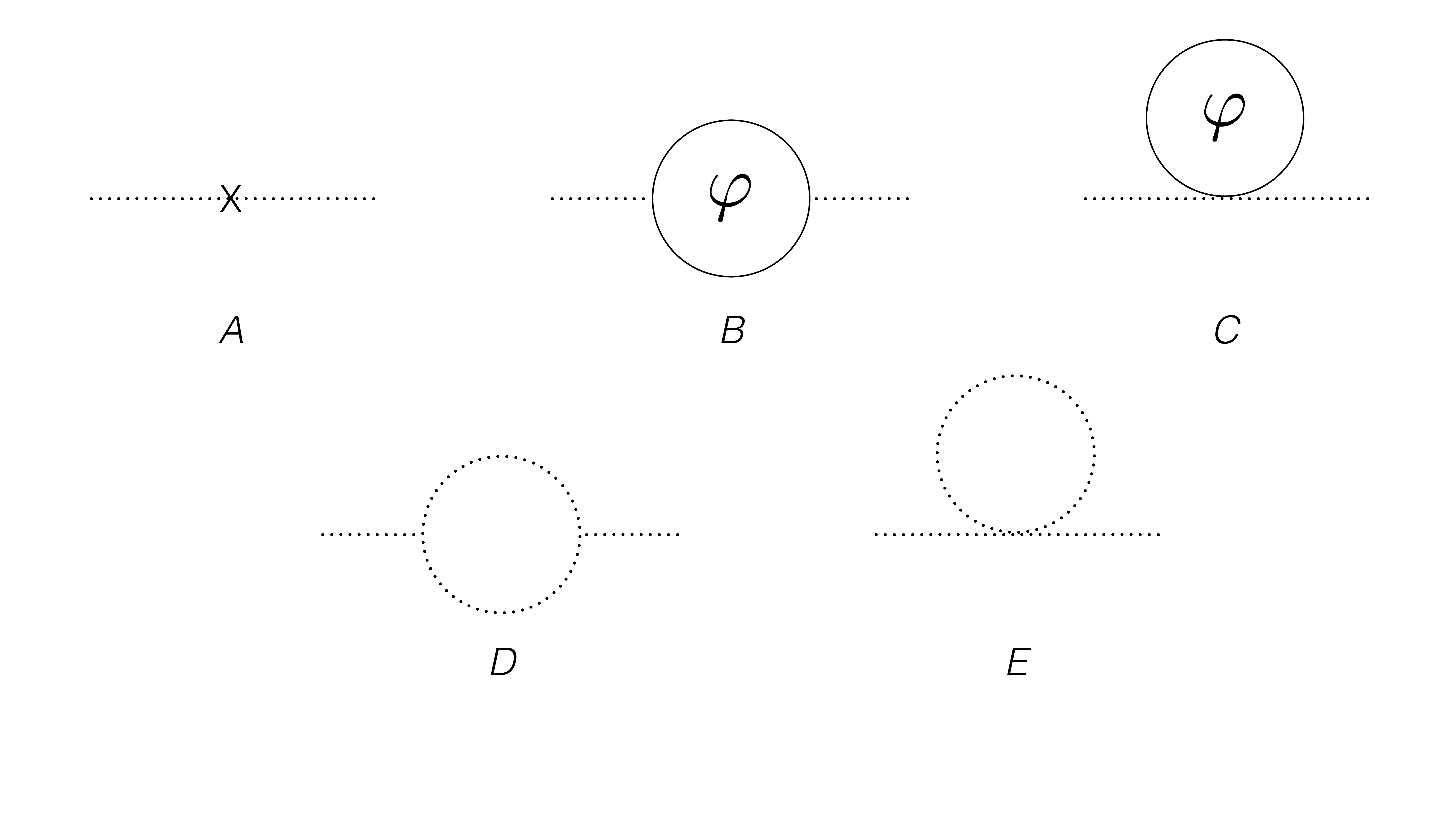}
\caption{One-loop radiative corrections to the graviton propagator, in QFT+GR. The solid and dotted lines denote matter and graviton propagators respectively.}
\label{figure_loop}
\end{center}
\end{figure}

Here, we argue that the standard renormalization of UV divergences in QFT does {\it not} affect the UV-IR coupling that we have discussed in this paper. 

The renormalization program in QFT (e.g., see \cite{WeinbergQFT}) accounts for the fact that the bare coupling constants that appear in the action are not directly observable. As such, the dependence on the UV cut-off that appears in radiative (or loop) quantum corrections to scattering amplitudes or propagators can be absorbed in a {\it renormalization} of the bare coupling constants of the theory.  

The radiative corrections to the graviton propagator, at one-loop order, are shown in Fig. (\ref{figure_loop}). Diagram A is the counter-term associated with the renormalization of the bare cosmological constant and Planck mass. Diagrams D and E are standard one-loop corrections to general relativity, which e.g., lead to higher order corrections to the Newtonian potential (e.g., \cite{Donoghue:1995cz}). Diagram C has the same exact structure as A, and thus can be absorbed into the renormalization of the Planck mass and cosmological constant (up to naturalness considerations). 

Diagram B is exactly what we have computed in Section \ref{sec:scalarfield}, and is the only non-trivial one-loop correction to the connected part of the metric two-point correlation function, due its interaction with a massive field. Furthermore, we saw that it is {\it finite}, and thus does not require renormalization. Another way to see this is to note that tree-level counter-terms renormalize the one-loop radiative corrections. However, in GR, scalar and vector modes are not dynamical, and thus their propagator vanishes at tree level. Therefore, the contribution of Diagram B to scalar and vector two-point correlation functions, which we have focused on in this paper, cannot be renormalized by GR local counter-terms (or otherwise require introducing non-local terms into the GR action).

\subsection{UV-IR Coupling in Gravity, Electromagnetism, and Chromodynamics}
\label{sec:EM}

Our study of the effect of stress-energy sources on the gravitational potential can be compared to the analogous case in quantum electrodynamics (QED), where charge and current density $J_\mu$ source the vector potential $A_\mu$ in Maxwell's equation. The conservation law $\partial_\mu J^\mu=0$ again constrains the two-point source correlator to have a tensor structure $\langle J_\mu J_\nu\rangle\propto\eta_{\mu\nu}-k_\mu k_\nu/k^2$ in momentum space, suppressing charge density fluctuations in the IR. Current fluctuations may have a flat Poisson spectrum, $\langle J_i(\k) J_j(-\k)\rangle\propto \delta_{ij}$, which acts as a source to the wave equation for the vector potential $A_i$ in the Coulomb gauge ($\partial_iA^i =0$). This is analogous to what happens to the tensor modes in GR, which also respond to their source through a hyperbolic wave equation. In either case, in contrast to scalar and vector modes that satisfy an elliptic equation, the response is an integral over the 4-volume of the past light-cone, rather than a spatial 3-hypersurface. This leads to additional cancellations due to fast oscillations in time and transverse (traceless) conditions, bringing down the degree of divergence from $\Lambda^5/k$ to $k^4 \log(\Lambda)$ for tensor modes, or $k^2 \log(\Lambda)$ for a vector potential. While the latter is already included in the standard treatments of divergences in QED (e.g., \cite{WeinbergQFT}), the CnC problem indicates a richer structure of IR divergence in massless spin-2 interacting theories, which has been so far overlooked. The infrared divergences of scalar and vector metric perturbations studied here are also different from infrared divergences due to soft gravitons in Feynman diagrams. The second is a dynamical effect which, like IR divergences in QED, is consistent with finite cross sections and observable rates \cite{Weinberg:1965nx,WeinbergQFT}, while the first is a non-dynamical consequence of the constraint equations, which cannot be treated in the same way.

A closer analogy to the CnC problem might be the confinement in quantum chromodynamics (QCD). As in the case of the CnC problem, the quantum fluctuations of the gluon field around its free vacuum blow up at an IR scale of $\Lambda_{\rm QCD}$.  Below this scale, the gluon field becomes strongly coupled, and an effective weakly coupled description is only possible in terms of different degrees of freedom (i.e. hadrons). In the case of gravity, this may suggest a non-geometric description on very large scales. However, we caution that this analogy must not be pushed too far, as the QCD IR divergence is still logarithmic, and not a power law as in the CnC (or CC) problem.

\subsection{The CnC Problem, CC Problem, and Effective Field Theory}

We emphasize that the cosmological non-constant problem as described here is complementary to, but distinct from, the cosmological constant problem. In both cases, a deeper understanding of UV physics is needed to explain the absence of any influence on geometry from the vacuum at high energies. However, different sign contributions to $T^{(V)}_{\mu\nu}$ both contribute positively to the two-point function, so vacuum-sourced fluctuations in geometry cannot cancel due to different sign contributions.

One may wonder why we could not just integrate out heavy fields in the IR, as is commonly done to find an effective field theory (EFT). Then, the lore is that UV physics will only renormalize the cosmological and Newton's constants, and otherwise has no observable consequence for low energy observables. In contrast, we have clearly demonstrated that the CnC problem has distinct consequences for cosmological observations that cannot be mimicked by a change in the cosmological or Newton's constants. 

To understand this better, we can look at the path integral description of (low energy) QFT+gravity:
\be
\int Dg D\varphi\times {\rm Diff}^{-1}[g,\varphi] \times \exp\left(i \int d^4x \sqrt{-g} \left\{R[g]+ {\cal L}_m[\varphi,g]\right\} \right). 
\ee
Here, ${\rm Diff}^{-1}[g,\varphi]$ is the measure of the path integral that effectively mods out the gauge degrees of freedom, in particular diffeomorphisms, in a relativistic theory. What is often called the gravitational EFT is given by integrating out the heavy matter fields, but ignoring the measure:
\be
\exp(i S_{\rm eff, naive}[g]) \equiv \exp(i S_{\rm GR}[g]) \times \int D\varphi \exp\left(i \int d^4x \sqrt{-g} {\cal L}_m[\varphi,g] \right),
\ee
which can be computed using, e.g., the heat kernel method as an expansion in powers of curvature invariants and their gradients (e.g., \cite{Vassilevich:2003xt}). However, in general
\be
{\rm Diff}^{-1}[g,0] \exp(i S_{\rm eff, naive }[g])  \neq  \exp(i S_{\rm GR}[g]) \times \int D\varphi \times{\rm Diff}^{-1}[g,\varphi]\times \exp\left(i \int d^4x \sqrt{-g} {\cal L}_m[\varphi,g] \right).
\ee
Therefore, the standard definition of EFT misses the path integral measure that is necessary for a sensible gravitational theory.  Admittedly, the non-perturbative measure of the gravitational path integral is unknown (although there is numerical evidence that a Lorentz violating preferred foliation might be necessary to yield physical results, e.g., \cite{Ambjorn:2013apa}). At the perturbative level we know that the role of the measure, along with the constraint equations, is to freeze out the non-dynamical degrees of freedom, i.e. scalar and vector metric perturbations, in terms of matter degrees of freedom. Therefore, in this admittedly non-covariant description, it is {\it not} possible to integrate out matter without integrating out part of the metric, which consequently leads to either a non-covariant or non-local effective action. In fact, this is exactly what we did in this in paper! Even though well-defined covariant measures exist in simpler gauge theories (such as Yang-Mills theory), to our knowledge, one does not exist for quantum gravity.    

From the point of view of canonical quantization, our result shows the non-local structure of the kinematic phase space (or Hilbert space) of QFT+gravity, imposed by elliptic GR constraint equations. At an intuitive level, this is similar to the Heisenberg uncertainty principle for momentum/position. However, in QFT+gravity the uncertainty is between the definition of the vacuum state (or QFT+gravity observables) in the UV and the IR.

In a broader context, the applicability of EFT in quantum gravity is already limited, due to the old cosmological constant problem, which suggests extreme fine-tuning of UV physics to explain our observable universe on large scales. On a more a theoretical ground, one may consider the {\it firewall paradox} \cite{Almheiri:2012rt,Braunstein:2014nwa} in the evaporation of black holes as another piece of evidence that local EFT in quantum gravity can fail even at very low energies, e.g., at the horizons of large black holes. Indeed, the non-locality introduced by imposing the GR constraint equations on the Hilbert space of quantum gravity, as discussed above, has been proposed as a possible resolution to the firewall paradox \cite{JACOBSON:2013ewa}.

\bigskip
{\bf Acknowledgements.}
We thank Cliff Burgess, Kurt Hinterbichler, Rafael Sorkin, and Richard Woodard for fruitful discussions, Andrew Tolley for suggesting the use of the K\"{a}ll\'{e}n-Lehmann representation, and Larry Ford for helpful correspondence in regards to stress-energy correlation functions. E.N. is supported in part by NSF Award PHY-1417385.  This work was partially supported by the Natural Science and Engineering Research Council of Canada, the University of Waterloo and by Perimeter Institute for Theoretical Physics. Research at Perimeter Institute is supported by the Government of Canada through Industry
Canada and by the Province of Ontario through the Ministry of Research \& Innovation. 

\bibliographystyle{JHEP}
\bibliography{CnCReferences}

\appendix

\section{K\"{a}ll\'{e}n-Lehmann Spectral Densities for Stress-Energy}
\label{app:KLrep}

In this Appendix we describe the tensor structures for stress-energy in the K\"{a}ll\'{e}n-Lehmann spectral representation, and give the spectral densities $\rho_{0,2}$ in terms of sums over states in Hilbert space.

Given the general form of Eq. \eqref{twopointKL}, we can extract $\rho_0$ or $\rho_2$ by contracting respectively with
\bea
S_0^{\mu\nu\alpha\beta} &\equiv& \frac{1}{9}g^{\mu\nu}g^{\alpha\beta}, \\
S_2^{\mu\nu\alpha\beta} &\equiv& \frac{1}{5}\left(\frac{1}{2}g^{\mu\alpha}g^{\nu\beta}+\frac{1}{2}g^{\mu\beta}g^{\nu\alpha} - \frac{1}{3}g^{\mu\nu}g^{\alpha\beta} \right),
\eea
since
\be
S_0^{\mu\nu\alpha\beta}\left(\frac{1}{2}P_{\mu\alpha}P_{\nu\beta}+\frac{1}{2}P_{\mu\beta}P_{\nu\alpha} - \frac{1}{3}P_{\mu\nu}P_{\alpha\beta}\right) = 0,
\ee
and
\be
S_2^{\mu\nu\alpha\beta}P_{\mu\nu}P_{\alpha\beta}=0.
\ee
Consequently, transforming Eq. \eqref{twopointKL} to momentum space, we find
\be\label{projectrho0}
S_0^{\mu\nu\alpha\beta} \langle T_{\mu\nu}(k)T_{\alpha\beta}(k')\rangle_{c,s} = (2\pi)^4 \delta^4(k+k')2\pi\rho_0(-k_4^2),
\ee
and from the second contraction,
\be\label{projectrho2}
S_2^{\mu\nu\alpha\beta} \langle T_{\mu\nu}(k)T_{\alpha\beta}(k')\rangle_{c,s} = (2\pi)^4 \delta^4(k+k')2\pi\rho_2(-k_4^2).
\ee
The spectral densities can be expressed in terms of a complete set of states,
\be
1=|0\rangle\langle0|+\sum_n\int\frac{d^4p}{(2\pi)^4} 2\pi\delta(p^2+M_n^2)\Theta(p_0) |\p,n\rangle\langle\p,n|.
\ee
Here, $\p$ is the total three-momentum for a given state $|\p,n\rangle$, and $M_n^2\equiv E_n^2 - \p^2$ is its invariant mass-squared, with the sum including continuous parameters such as relative momenta of particles. Inserting this sum between the operators in $\langle\hat{T}_{\mu\nu}(x)\hat{T}_{\alpha\beta}(y)\rangle_s$, and transforming to Fourier space as in Eqs. \eqref{projectrho0}-\eqref{projectrho2}, we identify the spectral density as
\be
\rho_0(\mu) = \frac{1}{2} \sum_n \delta(\mu-M_n^2) \left[ S_0^{\mu\nu\alpha\beta}\langle 0|T_{\mu\nu}(0)|\k,n\rangle\langle\k,n|T_{\alpha\beta}(0)|0\rangle + \ \rm c.c. \right],
\ee
with $\rho_2$ defined similarly with $S_2^{\mu\nu\alpha\beta}$. Only multi-particle states contribute to the sum over intermediate states. The contribution from the vacuum cancels when the disconnected part of the correlation function $\langle T_{\mu\nu}\rangle\langle T_{\alpha\beta}\rangle$ is subtracted off,
and $\langle 0|T_{\mu\nu}|1 \ \rm particle\rangle=0$ because $T_{\mu\nu}$ is quadratic in field operators and will only act on $\langle 0|$ to create states with two or more quanta.

\section{An Observational Limit on a Plateau of the CMB Anisotropy Angular Power Spectrum}
\label{SPT}

In Section \ref{sec:PoissonISW}, we showed that the Integrated Sachs-Wolfe (ISW) effect, caused by the metric perturbations that are sourced by vacuum fluctuations add a plateau to the power spectrum of CMB anisotropies $\Delta^2_l = \frac{l(l+1)C_l}{2\pi} $. Here, we derive an approximate upper limit on this plateau from current CMB observations. 

Given that $\Lambda$CDM predictions (due to Silk damping) as well as measurements of $\Delta^2_l$ are dropping rapidly at high $l$'s, the strictest bounds on a putative plateau would come from measurements at the highest $l$'s. The best current measurements are provided by the South Pole Telescope (SPT), at the highest $l$'s ($\lesssim$ 3000) \cite{Story:2012wx}. In order to estimate an upper limit on the size of an ISW plateau, we note that missing an additional term in the model causes a systematic increase in the $\chi^2$, given by:
\be
 \langle \chi^2_{\rm observed} \rangle \gtrsim \chi^2 +  \left[(\Delta^2_l)^{\rm ISW}\right]^2 \sum_{l} \frac{1} {\sigma^{2}_l}, \label{chi2}
\ee   
where the inequality is due to other potential systematic errors (further increasing $\chi^2$), while $\sigma_l$ is the SPT measurement error for  $\Delta^2_l$ in each l-bin. $\chi^2$ in Eq. (\ref{chi2}) is a random number which follows the $\chi^2$ distribution: 
\bea
P(\chi^2) d\chi^2 = \frac{\chi^{n-2} \exp\left(-\chi^2/2\right) d\chi^2}{2^{n/2} \Gamma(n/2)}, \\
n \equiv  [{\rm \# ~of~{\it l}-bins}]  -  [{\rm \# ~of~fitted~parameters}].
\eea
Note that we are also ignoring potential degeneracies between the ISW plateau and other cosmological parameters that are already fitted for. 

For the $\Lambda$CDM model best fit to SPT data $\chi^2_{\rm observed} = 45.9$, which uses 47 {\it l}-bins and 9 fitted parameters. Using the errors provided in Table 2 of \cite{Story:2012wx}, we can plug this into Eq. (\ref{chi2}), and then integrate the $\chi^2$ distribution to find:
\be
(\Delta^2_l)^{\rm ISW} < 2.2~ \mu K^2,
\ee
at 95\% confidence level. 

\end{document}